
\input phyzzx
\endpage
\centerline{Boris~ Spokoiny \dag}\footnote{\dag}
{Address after June 23,1993:
{}~Landau Institute for Theoretical Physics, Russian Academy of Sciences,
142432 Chernogolovka, Moscow region, Russia.}

\REF\Sta{A.A.Starobinsky, {\it Lecture Notes in Physics}.{\bf 246}
 (1986)107   \rm Proceedings
 of the seminar on  \it{Field Theory, Quantum Gravity
, and Strings},\rm Meudon
 and Paris VI,France 1984/85,edited by
 H.J.de Vega and N.Sanchez)
}
\REF\Gon{A.S.Goncharov, A.D.Linde and V.F.Mukhanov, Int.J.Mod.Phys.  A2
  (1987)  561.}

\REF\Sas{M.Sasaki, Y.Nambu \& K.Nakao,
{\it Nucl.Phys.}B{\bf308},  (1988)
 868;
  Phys.Lett.  B209 (1988) 197.}

\REF\Nam{Y.Nambu and M.Sasaki,Phys.Lett.B205 (1988) 441;
Phys.Lett.B219 (1989) 240.}

\REF\Nak{K.Nakao, Y.Nambu and M.Sasaki,Progr.Theor.Phys.80
(1988) 1041.}

\REF\Ort{A.Ortolan ,F.Lucchin \& S.Matarrese,
{ \it Phys.Rev.}D{\bf 38}(1988)465.
}
\REF\Hos{A.Hosoya, M.Morikawa and K.Nakayama,
Int.J.Mod.Phys.A4 (1989) 2613.}

\REF\Sal{D.S.Salopek \& J.R.Bond, {\it Phys.Rev.} D
{\bf 43}(1991) 1005.
}

\REF\Spok{B.L.Spokoiny, Deflationary Universe Scenario,
\nextline
Kyoto Univ.Preprint,May (1993)
}

\REF\Muk{V.F.Mukhanov,Zh.Eksp.Teor.Fiz.94 (1988) 1
  (Sov.Phys.  JETP  68  (1988)   1297);  Phys.Lett.  B218  (1989) 17. }

\REF\Mak{N.Makino \& M.Sasaki, {\it Progr.Theor.Phys.}
{\bf 86}(1991)103.}

\REF\Mukh{V.F.Mukhanov,Pis$^{,}$ma Zh.Eksp.Teor.Fiz.41 (1985) 402
 Sov.Phys.  JETP  Lett.  41   (1985)  493}

\REF\Luc{F.Lucchin \& S.Matarrese ,
{\it Phys.Rev.}D{\bf 32}(1985)1316;  Phys.Lett. 164B  (1985)  282.
}

\REF\Hal{J.Halliwell,{\it Phys.Lett.} B{\bf 185} (1987) 341.
}

\REF\Bar{J.D.Barrow,Phys.Lett.B187 (1987) 12.}

\REF\Yok{J.Yokoyama \& K.Maeda, {\it Phys.Lett.} B{\bf 207} (1988)31. }

\REF\Ste{D.H.Lyth and E.D.Stewart,  Phys.Lett.  B274 (1992) 168;
ibid.B302 (1993) 171.
}

\chapter{Introduction}

The stochastic approach to quasi-de Sitter Universes $[1-8]$ appeared to be
very
 Universes.
In a quasi-de Sitter inflationary model it is supposed
 that the evolution of the Universe is driven by some scalar
 field $\phi$ and the following
conditions which provide  slow rolling down of the scalar field
along the potential are satisfied:
$
\ddot{\phi} \ll H\dot{\phi},~~-\dot{H}/H^2 \ll 1.
$
It is easy to check that these conditions require the following restrictions
on the slope and curvature of the logarithm of the scalar field
potential
$ W(\phi)=logV(\phi)  :
$
$$
W'(\phi) \ll M^{-1}_{pl},
\eqno(1a)
$$
$$
W''(\phi) \ll M^{-2}_{pl}.
\eqno(1b)
$$
In this paper we develop a general technique
(good variables are found) for an arbitrary stochastic  superluminal
 evolution of the Universe
and
 {\it{explicitely}} generalize the stochastic approach to the models which do
{\
 the inflaton),
only the restriction (1b)
is supposed to be fulfilled.
We will call the potentials satisfying (1b) generalized exponential
 potentials since they
 include quasi-exponential potentials
$  logV(\phi)=\lambda\phi/M +log{\tilde V}(\phi)  $
with a slowly varying $  log{\tilde V}(\phi) $.
However maybe this is not a proper definition because in the case
$ \lambda =0 $
the potential does not look exponentially.
If
$ \lambda \not= 0 $
such potentials give rise to the (quasi) power-law evolution of the Universe
 which generally is not of the quasi-de Sitter form.
Up to now non de Sitter stochastic inflation was considered only for the case
of pure exponential potential (and pure power law evolution of the Universe)
[\Ort],[\Sal] and even this investigation seems to be incomplete for the
 following reason.
For the quasi-de Sitter
 Universe it is enough to consider only the perturbations of the scalar field
an
{\it pure}
 power-law inflationary Universe by introducing a new constant  (a diffusion
  coefficient) which was not calculated by them.
This is our first motivation for the detailed investigation of the models
 satisfying (1b).

Generalized exponential inflaton potentials
$$
V(\phi) = V_0 \exp{(-\lambda(\phi){\phi \over M})},
$$
where $M = M_{pl}/\sqrt{8\pi}$ is a reduced Planck mass,
with a slowly varying $\lambda(\phi)$
are interesting by themselves since they give us an opportunity to realize
an inflationary scenario without conversion of the false vacuum energy to
radiation[\Spok].
In this case inflaton is not \lq \lq frozen'' in the minimum of its
potential after the end of inflation and
its energy density may be still important at the present time giving us
dark matter and natural bias (since perturbations of the scalar field
decay after the wavelength became less than the size of the horizon).
There is also a natural mechanism that provides very small inflaton energy
density during the nucleosynthesis time which still allows to have
 (relatively) large inflaton energy density now.

The key idea of the scenario is that $\lambda(\phi)$ is not constant and
so the index $p = 2/\lambda^2 $  of the (quasi) power law evolution of the
Unive
first $a(t) \propto t^p$ with $p > 1$ (inflation) but then $a(t) \propto p'$
with $p' < 1/2 $(deflation) and later $p = 1/3$ (critical deflation).
There is particle production in an expanding Universe.
At the inflationary stage the energy density of the produced particles is
inflated out but the energy density of the particles produced after the
end of inflation when $p$ has already become less than 1 becomes more and
more dominant at the deflationary stage.
At the critical deflationary stage the Universe expansion is driven by the
inflaton kinetic energy term with the inflaton potential term rapidly
decaying.
Thus false - vacuum energy was (partly) converted to the inflaton kinetic
energy and partly dissipated due to Universe expansion.
At the critical deflationary stage radiation density produced after
thermalizati
of the created particles is small but growing and at last becomes dominated.
After the moment of time when radiation becomes dominant the kinetic energy
term starts to fall rapidly:$\dot{\phi}^2 \propto {1 \over a^6}$, the
 potential energy term is small since it was falling at the preceding
critical deflationary stage but it becomes almost frozen (since
$\dot{\phi} \propto {1 / a^3} \propto {1/ t^{3 \over 2}}$
is rapidly decaying).
So there is a big \lq \lq quasipure" radiation dominated epoch in this model
at which the contribution of the scalar field into the whole energy density
is
 negligible.
The nucleosynthesis period is supposed to be inside this epoch.
When the falling radiation energy density becomes of the order of (frozen)
inflaton potential energy density the Universe comes to the
 \lq \lq combined" stage at which the contribution of the inflaton energy
is comparable with that of ordinary matter.

In the ordinary inflationary model the random force in the equation for
the scalar field is effectively switched off after the inflationary stage
when the mass of the scalar field becomes much larger than the Hubble
parameter.
On the contrary in the case of the generalized exponential potential
the mass of the scalar field is always of the order of the Hubble parameter
or less.
So the random force is never switched off.
At the inflationary stage it is due to the modes going out of horizon
and at subsequent stages it is due to the modes coming inside horizon.
At the radiation dominated stage the classical motion of the inflaton
is very slow and the stochastic motion of the scalar field may
be even dominant.
This is the second motivation for our work.

\chapter{Stochastic dynamics in  general case
}

 We consider a model with one scalar field and gravity. The lagrangian of
 the theory is
 \vskip-9mm
     $$
 L=-{1\over2}M^2R +{1\over2}(\partial_{\mu}\phi)^2- V(\phi)
\eqno(2)
   $$
\
First we should solve the equations for a homogeneous background. In
a  synchronous coordinate system they read:
$$  3M^2 H^2 = {1\over2}\dot{\phi}^2 + V(\phi)~;~~~~~~~~~
\ddot{\phi} + 3H\dot{\phi} + V' = 0
 \eqno(2a)
$$
Then we may try to apply the stochastic approach to the model described by
(2). As usual [\Sta] we represent any operator $ F $ as a sum of a long wave
par
$ f(t,\vec r) $ (with a physical wave vector $ k_f = k/a < \epsilon H $)
and a short wave part (with $k/a > \epsilon H$ :
$$ F = f(t,\vec r) + \hat {I_\theta}(\hat{f}_s)
\eqno(3)
$$
where
$$ \hat{I_\theta}(\hat{f}_s)
 =  \int {d^3k\over {(2\pi)^{3\over 2}}}\theta(k- \epsilon aH)
{\hat f_s(t,k)}e^{i\vec k\vec r} ,
 \eqno (4)
 $$
$$ \hat{f}_s(t,k) = f_s(t,k)\hat{a}_k + {f_s}^*(t,k){\hat{a}_k}^+
\eqno(4a)
$$
and the perturbation $ f_s(t,k) $ is given in a synchronous coordinate system
and satisfies an exact wave equation for perturbations on the homogeneous
background satisfying eqs.(2a). (The subscript $ s $ shows that we use a
synchronous gauge.)
We have
$$ \phi = \varphi(t,\vec r) + \hat{I_\theta}(\hat\phi_s),
\eqno(5a)
$$
$$ \dot{\phi} = v(t,\vec r) + \hat{I_\theta}(\dot{\hat\phi_s}),
\eqno(5b)
$$
$$ \ddot\phi = w(t,\vec r) + \hat{I_\theta}(\ddot{\hat\phi_s}),
\eqno(5c)
$$
$$ g_{\alpha\beta} = g_{\alpha\beta}(t,\vec r) + \hat{I_\theta}
(\hat g_{\alpha\beta s}),
\eqno(5d)
$$
$$ \dot{g}_{\alpha\beta} = \kappa_{\alpha\beta}(t,\vec r) +
 \hat{I_\theta}(\dot{\hat g}_{\alpha\beta s}),
\eqno(5e)
$$
here $ g_{\alpha\beta} $ is a three dimensional space metric. We are to
substitute eqs.(5a)-(5e) into the equation for the scalar field and
Einstein's equations. The result may be represented in the form
$$
S_0 +S_1 +S_2 =0
\eqno(6)
$$
where $ S_0 $ is the contribution of the long wave part only, $ S_1 $
 is linear in the short wave part and $ S_2 $ is at least quadratic
 in the short wave
part.We find that $ S_1 = 0 $ since the short wave part perturbations
 satisfy the exact equations of motion for perturbations by definition.
We neglect higher order contributions,so we put $ S_2 =0 $ and we get
 $ S_0 = 0 $ . In eqs.
(5a)-(5e) we consider only the scalar mode of the system of metric and scalar
field perturbations. This allows us to consider
$$
{\kappa_\beta}^\alpha (t,\vec r) = 2H(t,\vec r){\delta_\beta}^\alpha
\eqno(7)
$$
with an accuracy$ \propto (k/aH)^2 \propto \epsilon^2 $ and in fact we
may put
$$
g_{\alpha \beta}(t,\vec r) = a^2(t,\vec r) \delta_{\alpha\beta}
\equiv e^{2\alpha(t,\vec r)} \delta_{\alpha\beta}
\eqno(8)
$$
Substituting (7) into the above equations $ S_0 = 0 $
we obtain with an accuracy $ \propto \epsilon^2 $
$$
 w + 3Hv + V'= 0,
\eqno(9)
$$
$$
H^2 = {{({{v^2}/2} + V(\varphi))}/ 3M^2}.
\eqno(10)
$$
Eq.(9) follows from the scalar field equation and eq.(10) is derived
from the $ _o^o $ -Einstein equation
$$
{1\over 8}[({\kappa_\alpha}^\alpha)^2 - {\kappa_\alpha}^\beta
{\kappa_\beta}^\alpha] +{1\over 2} R^{(3)} = [{1\over 2}(\dot{\varphi})^2 +
{1\over 2}(\nabla {\varphi})^2 + V(\varphi)]/M^2
\eqno(10a)
$$
where $ R^{(3)} $ is a three dimensional curvature and is neglected with
an accuracy $ \propto \epsilon^2. $

Differentiating (5a),(5b) and (5d) with respect to time and taking into
account (7)-(9) we find
$$
\dot{\varphi}(t,\vec r) = v(t,\vec r) + \hat{I}(\hat\phi_s),
\eqno(11)
$$
$$
\dot{v}(t,\vec r) = -3H(t,\vec r)v(t,\vec r) - V'(\varphi (t,\vec r))
+ \hat{I}(\dot{\hat{\varphi}_s}),
\eqno(12)
$$
$$
\dot{\alpha}(t,\vec r) = H(t,\vec r) + \hat{I}(\hat{\alpha_s}),
\eqno(13)
$$
where by definition
$$
 \hat{I}(\hat{f}_s)
 = \epsilon(aH)\dot{ } \int {d^3k\over {(2\pi)^{3\over 2}}}\delta(k- \epsilon
aH
{\hat f_s(t,k)}e^{i\vec k\vec r} ,
 \eqno (14)
$$
Now we are to consider the system (10)-(14). Deriving this system
we disregarded the terms of the second order in perturbations and
the terms of the order $ \epsilon^2 $ with respect to the main terms.
However we take into account all the terms of the first order. They
are contained in the random forces in eqs.(11)-(13). These equations
are in fact renormalization group equations for the long wave part
of the metric and the scalar field. As is well known, taking into
account the first order terms in a small parameter in renormalization group
equations corresponds to a special summation of an infinite number
of terms in the usual perturbation series.

Now we should try to solve the equations for perturbations on the
quasihomogeneous background satisfying (2a). It is more convenient
to do this in a longitudinal gauge where the metric is
$$
ds^2 = (1 - 2h)d{t'}^2 - (1+2h)a^2 dl^2
\eqno(15)
$$
and the action for the scalar perturbations (of the scalar field and
metric) is[\Muk - \Mak]
$$
  S=2M^4k^2\int d{\eta}[{\chi'_k}^2- (k^2 -  (U''/U) {{\chi}_k}^2)],
   \eqno(16) ,
 $$
where
$$
U= H/a\dot\varphi,
\eqno(16a)
$$
$ \eta $ is a conformal time and perturbations of metric and the scalar
field may be expressed in terms of $ \chi_k $ as follows:
$$
h_k = \dot{\varphi}\chi_k,
\eqno(17)
$$
$$
\delta\phi_k=-2M^2({\varphi''/ \varphi'a})\chi_k -2M^2({\chi'_k/ a}).
\eqno(18)
$$
We solve the equation
$$
\chi_k'' + (k^2 - U''/U)\chi_k = 0
\eqno(19)
$$
and normalize the solution so as
$$
\chi_k \rightarrow e^{-ik\eta}/(2M^2k\sqrt{2k})
\eqno(20)
$$
at $ k\eta \rightarrow - \infty $ (which corresponds to $ k/aH \rightarrow
\infty $ ). This normalization means that we have one degree of freedom for
the scalar mode as follows from eq.(16).
When $ k^2 \ll U''/U $ which roughly corresponds to the case $ k/a \ll H
$ we can easily solve the eq.(19). Two independent solutions $ c_1 U $
and $ c_2 U\int d\eta /U^2 $ after simple transformations taking into
account background equations (2a) give rise to [\Mukh]
$$
h_k = -A_k (1-{H\over a}\int ^{t'}a d\tilde{t}),
\eqno(21)
$$
$$
\delta\phi_k = {A_k\dot{\varphi}\over a}\int ^{t'}a d\tilde{t}.
\eqno(22)
$$
If we can solve eq.(19) we can find the constants $ A_k $.
After that we go back to the synchronous gauge.
For long wave perturbations (with $ k/a \ll H $ ) it is enough to make
a transformation of time :
$$
t = t' + T(t,\vec r),~~~~  ~~  ~~~T=\int dt h
\eqno(23)
$$
and perturbations of the scalar field and metric in the synchronous
coordinate system read:
$$
 \delta\phi_s(k,t) = \delta\phi_k + {\dot\varphi}T_k,
{}~~~~~                           \alpha_s(k,t) = h_k + HT_k ,
\eqno(24)
$$
where $ T_k $ is a Fourier component of $ T(t,\vec r) $.

So we should substitute perturbations (24) into our system (11)-(13).
However the obtained system is inconvenient because of the integral
term $ T_k $ in (24) (see (23)).
In fact our system appears to be a system
of integro-differential equations.
 Since it is inconvenient to study
this non-local system we try to change variables.

First we try to change gauge non-invariant perturbation of the scalar
field $ \phi_s(t,k) $ and its derivative $ \dot{\phi}_s (t,k) $
into
$$
 \xi(t,k) = \phi_s(t,k) - {(\dot\varphi/ H)}\alpha_s (t,k)
 = \delta\phi_k - {(\dot\varphi/ H)}h_k \approx A_k\dot{\varphi}/H
\eqno(25)
 $$
and
$$
 \psi(t,k) =\dot{ \phi}_s(t,k) - {(\dot v/ H)}\alpha_s (t,k)
 =\delta\dot{\phi}_k - (\ddot {\phi}/ H - \dot \varphi)h_k
 \approx A_k\ddot{\varphi}/H
\eqno(26)
 $$
correspondingly which do not contain the integral quantity $ T_k $.
This may be easily achieved by taking the logarithm of the scale
factor to be a new time variable.
In this case we obtain from (11)-(13):
$$
d\varphi/d\alpha = \vartheta + (1/H)\hat{I}(\hat{\xi}),
\eqno(27)
$$
$$
dv/d\alpha = -3v -V'/H + (1/H)\hat{I}(\hat{\psi}),
\eqno(28)
$$
where
$$
\vartheta = v/H,
\eqno(29)
$$
$ \hat{\xi} $ and $ \hat{\psi} $ are obtained from $ \xi(t,k) $ and
$ \psi(t,k) $ according to (4a).
We see that $ \xi(t,k) $ in (25) is a gauge invariant perturbation
so a logarithm of the scale factor seems to be a more fundamental
time variable than a synchronous time.

\chapter{Generalized power law evolution of the Universe}

We will be especially interested in the  quasipower-law
evolution of the Universe since in this case we may solve the
eq.(19) for perturbations explicitely. First we consider a pure
exponential potential
$$
V(\phi) = (1/2)V_0 \exp{(\lambda\phi/M)}
\eqno(30)
$$
The equations for the homogeneous background have a power-law solution
 [\Luc]
\vskip-7mm
 $$
a(t)=t^{n+1},~~~~ \phi=\varphi(t)=\varphi_0 - \sqrt{2(n+1)}M\log t,
\eqno(31)
$$
\vskip-3mm
\noindent
where $ n={2/{\lambda^2}}-1 $. Solution (31)is an asymptotically dominated
 solution of
the system (2a) (it was studied in [\Hal,\Bar, \Yok]).
In the solution (31) $ \vartheta $ is constant and this fact leads us
to the conclusion that it could be a more convenient variable than
$ v $.

Now we consider again a general potential satisfying (1b).
Instead of eq.(10) we will use
$$
H^2 = 2V/(6M^2 - \vartheta^2)
\eqno(32)
$$
Eq.(28) after taking into account eqs.(27) and (32) may be transformed
to
$$
 {d\vartheta \over d\alpha} = -(3-{\vartheta^2\over 2M^2})(\vartheta +
 M^2{V'\over V}) + \chi ,
 \eqno(33)
 $$
where
$$
\chi = {1\over H}\hat{I}(\hat{\Psi}),~~~~~
\hat{\Psi} = {V\over 3M^2 H^3}(\hat{\psi} -
 {\dot{\varphi}V'\over 2V}\hat{\xi})
\eqno(34)
$$
or taking into account (25)-(26)
$$
\Psi(t,k) \approx {A_k V\over 3M^2 H^4}
(\ddot{\varphi} - \dot{\varphi}^2{V'\over 2V}) =
-A_k {V\over M^2 H^2}(\vartheta + M^2(logV)') =
2A_k{d\vartheta\over d\alpha}.
\eqno(35)
$$
Since $ \vartheta $ in the background solution (31) is constant we obtain
$$
\Psi(t,k) \approx 0.
\eqno(36)
$$
So $ \chi=0 $ in the limit $ \epsilon \rightarrow 0 $!
This is a very important result since it shows that we have chosen
good variables ($ \vartheta $ and $ \alpha $).
In the derivation of the eq.(36) we used a particular
model with a pure exponential potential.
Now we would like to extend our model and consider quasi-exponential
potentials.
So we should check the validity of (36) again.
The system of background equations (2a) is equivalent to two equations
(27) and (33) without random forces,eq.(27)being just the definition
of $ \vartheta $.
{}From eq.(33) in the first approximation we obtain
$$
 \vartheta = -M^2(logV)' \equiv -M^2 W'.
 \eqno(37)
 $$
In the second approximation we get
$$
\vartheta = -M^2 W'[1+M^2 W''/(3 - M^2 {W'}^2 /2)].
\eqno(38)
$$
For the correction to eq.(37) to be small we need for the requirement
(1b) to be fulfilled.
Substituting the approximate solution (38) into (35) we obtain
$$
\Psi(t,k) \approx A_k M^4 W'W''
\eqno(39)
$$
and from (25)
$$
\xi(t,k) \approx A_k \vartheta \approx -A_k M^2 W'
\eqno(40)
$$
So
$$
-\Psi(t,k)/\xi(t,k) \approx -{1\over \vartheta}{d\vartheta\over d\alpha}
\approx M^2 W'' \ll 1
\eqno(41)
$$
and we may neglect $ \Psi $ with respect to $ \xi $.
Then the solution to eq.(33) is given by (37).
A general solution to the eq.(33) contains two modes.
A slowly varying mode (37) is an asymptotically leading mode.
The other mode decays as
$ exp{(-(3 - \vartheta^2 /2M^2 )\alpha)} $.
Substituting (37) into (27) we obtain the Langevin equation
$$
 {d\varphi/ d\alpha} = -M^2{V'/ V} + \eta(\alpha),
\eqno(42)
$$
where $ \eta(\alpha) $ is a white noise with the correlation function
$$
\VEV {\eta(\alpha)\eta({\alpha'})} =
{1\over 2\pi^2}(1 + {\dot{H}\over H^2})
(k^3 \vert \xi_k \vert ^2 \vert _{k = \epsilon aH}
\delta(\alpha - \alpha'),
\eqno(43)
$$

Eq.(19) on the background (31) may be easily solved in Bessel functions.
\footnote{*}{See also independent calculation [\Ste] in another context.}
 The solution that has a proper asymptotic behavior (20) at $ -k\eta \gg 1 $
 is of the form:
                                             \vskip-5mm
$$
 \chi_k = - {\sqrt\pi\over 2}{1\over 2kM^2}\sqrt{-\eta}
H^{(1)}_{{1\over 2}+ {1\over n}}(-k\eta)
\eqno(44)
$$
The perturbation of the scalar field is given by (18) and at $ -k\eta \ll 1 $
we find
$$
 \delta\phi_k = -{i\over {\sqrt2}}{H_*^{1+{1\over n}}
\over k^{{3\over 2} + {1\over n}}}F(n)
\eqno(45)
$$
where $ H_* $ is defined by the equation for the Hubble parameter:
\vskip-5mm
$$
 H(\eta) = (1+{1\over n})H_*(-H_*\eta)^{1\over n},
\eqno (46)
$$
$$
      F(n)= {2^{1\over n}\sqrt{\pi}(1 + {1\over n})\over
 cos{\pi\over n}\Gamma({1\over 2}-{1\over n})}
 \eqno(47)
 $$
( For the de Sitter space $ F=F(\infty)=1$). The perturbation of the metric
 at $ -k\eta\ll1 $ is
 $$
 h_k = -{iF(n)\over 2\sqrt{n+1}}{H_*^{1+{1\over n}}
\over Mk^{{3\over 2}+{1\over n}}}.
 \eqno(48)
 $$
Now we should substitute (45) and (48) into (25)-(26) and then into (43).
 Of course we take only the leading terms in a series in $ (-k\eta) $ to obtain
$
 \delta \propto max(\epsilon^2,{1\over n} \epsilon^{1 + {2\over n}})
 $
due to $ \delta(k - \epsilon aH) $. So we may neglect this contribution to $
\ch
So we put $ \chi = 0 $.
Eq.(43) gives us
$$
\VEV {\eta(\alpha)\eta({\alpha'})} = 2{D_0(\varphi)}{V(\varphi)\over M^2}
{1-{M^2\over 2}{({V'\over V})^2}\over {1 - {M^2\over 6}{({V'\over V})^2}}}
\delta(\alpha - \alpha'),
\eqno(49)
$$
where
$$  n\equiv n(\varphi) = 2({V\over MV'})^2 - 1,~~~~
 2D_0(\varphi) = {1\over 12{\pi}^2}{1\over \epsilon^{2\over n}}{(n + 2)^2\over
(
 \eqno(50)
  $$
 From the Langevin eq.(42) we obtain a Fokker-Planck equation
\nextline
 (using the
Stratonovich formulation)
$$
 {\partial P\over \partial \alpha} = M^2 {\partial \over \partial
\varphi}({V'\o
 \eqno(51)
  $$
where
$$
 {A(\varphi)}^2 =D_0(\varphi){1 - {M^2\over 2}({V'\over V})^2\over {1 -
{M^2\ove
 \eqno(52)
  $$

As usual the Fokker-Planck equation (51) has two stationary solutions,one of
whi
$$
 P_0(\varphi) = {const\over A(\varphi)V^{1\over 2}}\exp[M^4\int {1\over
{A(\varp
\eqno(53)
 $$
corresponds to the "quantum creation" of the Universe since it has the
Hawking-M
$$ P(\varphi)\approx {j_0V/ M^2V'}
\eqno(54)
$$
and describes the "classical creation" of the Universe,i.e.,the evolution from
t

\chapter{Approximate solutions to the Fokker-Planck equation}

We can also find an approximate solution to the Fokker-Planck equation (51)
with a delta-function initial condition :
$$
P(\varphi, \alpha_{in}) = \delta (\varphi - \varphi_{in}).
$$
It is convenient to change variables:
$$
\int^{\varphi}  _{\varphi_{in}} d\varphi '/\sqrt{V(\varphi ')}  =  \Phi,
{}~~~~~~~~~\sqrt{V(\varphi)} P = Q.
\eqno(55)
$$
We restrict ourselves to the case
$$
W''/ W' \ll W'
\eqno(56)
$$
since in this case we can explicitely  perform the integration in (55).
We obtain
$$
\Phi \simeq -{2 \over W'(\varphi) \sqrt{V(\varphi)}}
\eqno(57)
$$
and the Fokker-Planck equation (51) gets a quasi-linear form:
$$
 {\partial Q\over \partial \alpha} = -{1 \over 2}M^2
 {\partial \over \partial \Phi}(W'^2(\Phi)  \Phi Q) +
 {1\over M^2}{\partial\over \partial \Phi}[A(\Phi)
 {\partial\over \partial \Phi}A(\Phi)Q],
 \eqno(58)
  $$
since if
$ n $ or $ 1/ \log{1 \over \epsilon} $ are not very small
$$
A'(\Phi)/A(\Phi) \ll W'
\eqno(59)
$$
as a consequence of eq.(56).
Taking into account (56) and (59) we see that
$ W'(\varphi) $ and $ A(\varphi) $ are slowly varying quantities and
 we may substitute
$ \varphi = \varphi_{cl} (\alpha) \equiv \varphi_{cl}$ into them.
Here
$ \varphi_{cl}(\alpha) $
is the solution of the equation
$$
 {d\varphi/ d\alpha} = -M^2 W'
\eqno(60)
$$
with the initial condition
$ \varphi(\alpha_{in}) = \varphi_{in}.$

Now we would like to remind that the solution of the equation
$$
 {\partial P\over \partial t} +f(t){\partial \over \partial x}(x P) =
{1 \over 2}{\partial^2 P \over \partial^2 x}
 \eqno(61)
  $$
with the initial condition
$ P(x,t_{in}) = \delta (x - x_{in}) $
is
$$
P(x,t) = {1 \over {\sqrt{2\pi \Delta_0 (t)}}}
\exp{[-{(x - x_0 (t))^2 \over \Delta_0 (t)}]}
\eqno(62)
$$
where
$$
x_0 (t) = x_{in} \exp{(f(t) - f(t_{in}))},~~~
\Delta_0 (t) = \int^t _{t_{in}} D(t') \exp{(2f(t) - 2f(t'))} dt'.
\eqno(63)
$$
One may easily obtain (62) by making a Fourier transform in $ x $
of  eq.(61).
The correspondence between (61) and (58) is the following:
$$
f(\alpha) = -W(\varphi_{cl})/2,~~~~~
D(\alpha) = 2A^2 (\varphi_{cl}) / M^2,
\eqno(64)
$$
so we obtain
$$
P(\varphi,\alpha) = {1 \over \sqrt{2\pi V(\varphi)}}
\exp{[-{( \Phi(\varphi) \sqrt{V(\varphi_{cl})} -
\Phi(\varphi_{in}) \sqrt{V(\varphi_{in})})^2 \over 2\Delta(\alpha)}]}
\eqno(65)
$$
where $ \Phi(\varphi) $ is given by (57) and
$$
\Delta (\alpha) = \tilde{\Delta}(\varphi_{in}) -
 \tilde{\Delta}(\varphi_{cl}),~~~~~~~
\tilde{\Delta}(\varphi) =2{A^2 (\varphi) \over W'^2 (\varphi)}
 {V(\varphi) \over M^4},
\eqno(66)
$$
generalizing the result of ref.[\Sal] obtained for the case of pure
exponential potential.
Consideration of the boundary effects (due to the boundary condition
at $\varphi = \varphi_0$,where $V(\varphi_0) \sim M^4 _{pl}$ ) is
qualitatively similar to the case of pure exponential potential that
is done in [\Sal].

\centerline{Acknowledgements}

I am very grateful to Prof.M.Sasaki for many very useful discussions on
this subject, to Dr.Ewan Stewart for critical remarks  and to Prof.H.Sato
 for the elucidating question on the
asymptotic nature of the considered solutions asked at the Astrophysics
Group Colloquim in Kyoto University.
This work was supported in part by the JSPS  Fellowship and by
Monbusho Grant-in-Aid for Encouragement
of Young Scientists,No.92010.

\refout

\end